\def\year{2023}\relax
\documentclass[letterpaper]{article} % DO NOT CHANGE THIS
\usepackage{aaai23}  % DO NOT CHANGE THIS
\usepackage{times}  % DO NOT CHANGE THIS
\usepackage{helvet}  % DO NOT CHANGE THIS
\usepackage{courier}  % DO NOT CHANGE THIS
\usepackage[hyphens]{url}  % DO NOT CHANGE THIS
\usepackage{graphicx} % DO NOT CHANGE THIS
\usepackage{natbib}  % DO NOT CHANGE THIS AND DO NOT ADD ANY OPTIONS TO IT
\usepackage{caption} % DO NOT CHANGE THIS AND DO NOT ADD ANY OPTIONS TO IT
\DeclareCaptionStyle{ruled}{labelfont=normalfont,labelsep=colon,strut=off} % DO NOT CHANGE THIS
\usepackage{algorithmic}
\usepackage{newfloat}
\usepackage{listings}
\usepackage{bm}
\usepackage{subcaption}
\usepackage{amsmath}
\usepackage{booktabs}
\usepackage{xcolor}
\usepackage{threeparttable}
\usepackage{amsthm}
\usepackage{amsfonts}
\usepackage{mathtools}
\usepackage[ruled,linesnumbered]{algorithm2e}

\let\oldnl\nl% Store \nl in \oldnl
\newcommand{\nonl}{\renewcommand{\nl}{\let\nl\oldnl}}% Remove line number for one line
\SetKwInput{KwData}{Input}
\SetKwInput{KwResult}{Output}
\DontPrintSemicolon

\title{Communication-Efficient Search under Fully Homomorphic Encryption for Federated Machine Learning}
\author{
Dongfang Zhao\\
\textit{University of Washington}\\
dzhao@uw.edu
}

\begin{document}

\maketitle
\thispagestyle{plain}
\pagestyle{plain}

\begin{abstract}
Homomorphic encryption (HE) has found extensive utilization in federated learning (FL) systems, capitalizing on its dual advantages: (i) ensuring the confidentiality of shared models contributed by participating entities, and (ii) enabling algebraic operations directly on ciphertexts representing encrypted models. Particularly, the approximate fully homomorphic encryption (FHE) scheme, known as CKKS, has emerged as the de facto encryption scheme, notably supporting decimal numbers. While recent research predominantly focuses on enhancing CKKS's encryption rate and evaluation speed in the context of FL, the search operation has been relatively disregarded due to the tendency of some applications to discard intermediate encrypted models. Yet, emerging studies emphasize the importance of managing and searching intermediate models for specific applications like large-scale scientific computing, necessitating robust data provenance and auditing support. To address this, our paper introduces an innovative approach that efficiently searches for a target encrypted value, incurring only a logarithmic number of network interactions. The proposed method capitalizes on CKKS's additive and multiplicative properties on encrypted models, propagating equality comparisons between values through a balanced binary tree structure to ultimately reach a single aggregate. A comprehensive analysis of the proposed algorithm underscores its potential to significantly broaden FL's applicability and impact.
\end{abstract}

\section{Introduction}

Homomorphic encryption (HE) has been widely employed in federated learning (FL)~\cite{eaha_ppet23}, offering a groundbreaking solution to the critical challenge of preserving data privacy while enabling collaborative machine learning. By allowing computations to be performed on encrypted data without the need for decryption, HE maintains the confidentiality of individual data contributions throughout the federated learning process. This cryptographic technique has gained prominence due to its ability to facilitate the secure aggregation of model updates from decentralized devices or servers, thereby ensuring that sensitive information remains protected. As federated learning continues to emerge as a powerful paradigm for training machine learning models across distributed environments, the integration of HE not only safeguards the privacy of participants but also paves the way for efficient and privacy-preserving collaborative AI advancements.

Due to the nature of Federated Learning (FL) applications, the data types primarily consist of decimal values. This prevalence of decimal data is a result of the diverse range of domains that utilize FL, spanning areas such as healthcare, finance, and IoT. The use of decimal values is often driven by the need to capture and process nuanced information with high precision, as is the case with medical measurements, financial transactions, and sensor readings. In FL, where machine learning models are collaboratively trained across decentralized devices or servers while keeping data localized, the representation of decimal values becomes essential for maintaining the integrity and accuracy of the information being shared. Consequently, addressing the challenges and intricacies associated with handling decimal data within the FL framework becomes a crucial endeavor, influencing the design of secure and efficient protocols for privacy-preserving federated learning.

Because CKKS~\cite{ckks} (Cheon-Kim-Kim-Song) remains the only Homomorphic Encryption (HE) scheme supporting decimal values, recent research efforts have been concentrated on optimizing CKKS for various applications. Specifically, these endeavors revolve around enhancing two pivotal aspects: the encryption of model updates and the aggregation of local models. CKKS's unique capability to operate on real numbers directly makes it a vital candidate for preserving the precision and accuracy required by decimal-based datasets, commonly encountered in domains like finance, scientific computing, and cryptography. As a result, refining the encryption techniques for model updates ensures that the privacy of sensitive data is upheld during collaborative training in Federated Learning (FL). Additionally, focusing on the aggregation of local models under the constraints of CKKS aims to guarantee efficient and accurate fusion of contributions from various devices or servers in the FL paradigm. This concerted research in CKKS optimization not only advances the capabilities of HE for real-world applications but also contributes to the continued evolution of privacy-preserving technologies in the era of decentralized and collaborative computing.

Although encryption and evaluation operations have been extensively studied in the realm of Homomorphic Encryption (HE), there remains a notable gap in research pertaining to other equally crucial operations, particularly the search operation on CKKS ciphertexts. While much attention has been directed toward enabling computations on encrypted data, the unique challenges and intricacies associated with searching encrypted data have received relatively limited exploration. This discrepancy in focus is particularly significant considering the pivotal role that search operations play in various applications, such as scientific computing, where data provenance and auditing are of paramount importance. In fields like scientific research, where maintaining the integrity and origin of data is critical, the ability to search encrypted data in a privacy-preserving manner offers a compelling avenue for advancing security and accountability. As CKKS is a promising HE scheme known for its support of real number arithmetic and precision, exploring search operations within this framework has the potential to reshape the landscape of secure data management and utilization, leading to novel solutions that align with the demands of modern decentralized and privacy-conscious computing paradigms.

This paper proposes a worst-case optimal algorithm for searching in the CKKS ciphertexts.
Specifically, the proposed algorithm incurs only a logarithmic number of interactions between the client and the server. 
Our algorithm does not require the expensive preprocessing, e.g., sorting, over the ciphertext on the server,
which takes $\mathcal{O}(n \log n)$;
rather, the server only applies a linear number of algebraic operations over the ciphertexts.
The key idea is to leverage the additive and multiplicative homomorphism in CKKS and construct a balanced binary tree where each intermediate node stores some compaction information in a fashion similar to a Merkle tree~\cite{merkletree}.

We start with some preliminaries and related work.

\section{Preliminaries and Related Work}

\subsection{Federated Learning}

% The term federated learning (FL) was first coined in~\cite{mcmahan_aistat17} to allow clients to contribute to distributed machine learning without disclosing their private data.
% As FL continues to be more preferred compared to the conventional machine learning approaches, it still has numerous areas of improvement that researchers continuously seek to address. 
% For example, authors of~\cite{cui2021} proposed a method called Fair and Consistent Federated Learning (FCFL) to tackle disparity and performance inconsistency in FL. 
% In~\cite{gong2022}, authors proposed the use of ensemble cross-domain knowledge distillation to preserve privacy in FL. 
% Authors of~\cite{hamer2020} utilized ensemble algorithms to alleviate the communication cost of both server-to-client and client-to-server communication in FL. 
% In~\cite{Reisizadeh2020}, authors proposed a method called FedPAQ to address some of the communication and scalability challenges of FL.

The term federated learning (FL) was initially coined in the work by McMahan et al.~\cite{mcmahan_aistat17}, aiming to facilitate distributed machine learning where clients can contribute without compromising their private data. While FL has gained substantial preference over conventional machine learning approaches, there remain several areas for improvement that researchers consistently strive to address. For instance, Cui et al.~\cite{cui2021} introduced Fair and Consistent Federated Learning (FCFL), a method designed to mitigate disparities and performance inconsistencies in FL. In a similar vein, Gong et al.~\cite{gong2022} proposed the application of ensemble cross-domain knowledge distillation to safeguard privacy within the FL framework. Hamer et al.~\cite{hamer2020} leveraged ensemble algorithms to alleviate communication costs in both server-to-client and client-to-server interactions in FL. Furthermore, Reisizadeh et al.~\cite{Reisizadeh2020} presented FedPAQ, a method addressing communication and scalability challenges within the FL paradigm. As FL remains a dynamic field, these innovative approaches collectively contribute to its ongoing refinement and optimization.

\subsection{Attacks on Federated Learning}

% A poisoning attack is one of the major forms of attack that could arise while using machine learning models. 
% For example, when adversary or malicious clients deliberately add compromised samples to the training pool of the model~\cite{alfeld2016},
% it was called a \textit{data poisoning} attack. 
% Similarly, a malicious client can upload an arbitrary model to the aggregator---a \textit{model poisoning} attack (also termed adversarial attack)~\cite{li2020review}.
% Many known attacks belong to this category, such as backdoor attacks~\cite{yang2019, bhagoji2019analyzing, BagdasaryanVHES20, Xie2020DBA}. 
% Several defense approaches have been proposed to alleviate poisoning attacks in FL. 
% These approaches can be broadly classified as server-based or client-based defense approaches. 
% Sun et al.~\cite{sun2021} proposed a client-based defense, named White Blood Cell for Federated Learning (FL-WBC) to mitigate model poisoning attacks that have already polluted the global model. 
% Yin et al.~\cite{yin2018} proposed a server-based defense approach through robust aggregation to improve the robustness of FL against model poisoning attacks. 
% Similarly, Sun et al.~\cite{sun2019} proposed a server-based defense by clipping local updates to mitigate poisoning attacks. Krum~\cite{pblan_nips17} mitigated poisoning attacks in FL through the utilization of the similarity of benign clients’ local updates. 
% Shejwalkar and Houmansadr~\cite{Shejwalkar2021} designed a defense against FL poisoning called divide-and-conquer.

The threat of a poisoning attack looms large as one of the primary risks in the realm of machine learning models. For instance, when adversaries or malicious clients deliberately inject compromised samples into the model's training dataset~\cite{alfeld2016}, this form of attack is referred to as a \textit{data poisoning} attack. In parallel, a malicious client could also upload a tampered model to the aggregator, constituting a \textit{model poisoning} attack (also known as an adversarial attack)~\cite{li2020review}. This category encompasses numerous well-documented attacks, including the likes of backdoor attacks~\cite{yang2019, bhagoji2019analyzing, BagdasaryanVHES20, Xie2020DBA}. To counteract these threats, various defense strategies have emerged within the realm of Federated Learning (FL), broadly categorized as either server-based or client-based approaches. Sun et al.~\cite{sun2021} introduced a client-based defense coined as White Blood Cell for Federated Learning (FL-WBC), aimed at mitigating model poisoning attacks that have infiltrated the global model. In a server-based defense, Yin et al.~\cite{yin2018} proposed robust aggregation to bolster FL's resilience against model poisoning attacks, while Sun et al.~\cite{sun2019} employed local update clipping for similar purposes. Krum~\cite{pblan_nips17} harnessed the similarity among benign clients' local updates to fend off poisoning attacks in FL. Additionally, Shejwalkar and Houmansadr~\cite{Shejwalkar2021} devised a defense strategy named divide-and-conquer to counter FL poisoning threats effectively. These pioneering endeavors collectively contribute to the development of a fortified FL landscape, with strategies aimed at safeguarding against the ever-evolving spectrum of poisoning attacks.

\subsection{Homomorphic Encryption}

Homomorphic encryption, a groundbreaking cryptographic technique, empowers computations on encrypted data while upholding the privacy of the underlying information. An encryption function $f()$ exhibits the trait of being \textit{additive homomorphic} when it satisfies the equation $f^{-1}\left(f(a) \oplus f(b)\right) = a + b$, with $f^{-1}()$ signifying the decryption function and $\oplus$ representing the binary operation within the scope of $f()$. The concept of fully homomorphic encryption (FHE) broadens this attribute to encompass both additive and multiplicative operations.
Although diverse FHE schemes, including BFV~\cite{bfv} and CKKS~\cite{ckks}, have been conceived, their integration into Homomorphic Encryption for Federated Learning remains limited due to their pronounced computational overhead. Instead, HE schemes that exclusively support additive or multiplicative homomorphism exhibit greater efficiency and have thus found their way into the fold of Privacy-Preserving Machine Learning (PPML) systems~\cite{symmetria_vldb20}. A pertinent example is the Paillier~\cite{ppail_eurocrypt99} encryption scheme, which has been harnessed in endeavors like~\cite{shardy_arxiv17,zhang_atc20} to ensure privacy preservation within the context of federated learning.

\subsection{CKKS Scheme}

The CKKS~\cite{ckks} homomorphic encryption scheme represents a transformative advancement in the field of cryptography, offering a potent solution for secure computation on encrypted data while accommodating real-number arithmetic. At its heart, CKKS employs a unique representation of ciphertext as a pair of polynomials, strategically leveraging mathematical operations on these polynomials to facilitate computations on encrypted values. This distinctive approach not only preserves the precision of real-number data but also ensures the confidentiality of sensitive information. CKKS's dual capability to perform both additive and multiplicative operations on encrypted data renders it highly adept for applications requiring precise computations on private data.
However, the realization of these benefits comes with certain challenges. The arithmetic operations in CKKS tend to enlarge ciphertext and introduce noise, necessitating countermeasures such as \textit{relinearization} to manage the growth in ciphertext size. This process optimizes the performance of multiplication operations while maintaining manageable ciphertext dimensions. Additionally, the increased noise due to homomorphic operations necessitates "modulo switching" to manage noise accumulation. This procedure strategically refreshes the ciphertext noise levels, allowing secure and accurate computations while safeguarding data privacy.

\subsection{Security Proof}

The security of cryptographic schemes, including homomorphic encryption, is often assessed through formal proofs based on established security notions. The Indistinguishability under Chosen Plaintext Attack (IND-CPA) security is a fundamental concept that assures the confidentiality of encrypted data even when adversaries have access to chosen plaintext-ciphertext pairs. The proof of IND-CPA involves the application of reduction techniques, analysis of probabilities, and the concept of negligible functions.

\paragraph{Reduction Techniques}
The proof template for IND-CPA employs reduction, where the security of the cryptographic scheme is reduced to the assumed infeasibility of a particular computational problem. This is usually the hardness of breaking the underlying cryptographic primitive. In the context of homomorphic encryption, this reduction demonstrates that if an adversary can distinguish between two ciphertexts encrypted from distinct plaintexts, then the adversary could also solve the assumed computational problem, which is considered hard.

\paragraph{Probabilistic Analysis}
IND-CPA security hinges on the concept that an adversary should not be able to distinguish between two ciphertexts that correspond to different plaintexts. The proof template involves a probabilistic analysis, where the probability that an adversary can distinguish between these ciphertexts is shown to be negligibly close to 0. This probabilistic approach ensures that, even with an unbounded adversary, the probability of successful distinguishing remains negligible.

\paragraph{Negligible Functions}
Central to the proof template are negligible functions, which are functions that decrease faster than the reciprocal of any polynomial. In the context of security proofs, if the advantage of an adversary (the probability that they can distinguish between ciphertexts) is bounded by a negligible function, then the scheme is considered IND-CPA secure. This notion encapsulates the concept that as the input size grows, the negligible function diminishes more rapidly than any polynomial increase.

\section{Search over CKKS Ciphertexts}

\subsection{Notations}
Fully Homomorphic Encryption (FHE) introduces a set of notations that are commonly used to describe its operations and properties. Let $Enc$ denote the encryption function that takes a plaintext $m$ and transforms it into a ciphertext $c$ using the public key, i.e., $c = Enc(m)$. The decryption function, $Dec$, inversely converts the ciphertext back to the plaintext using the secret key, yielding $m = Dec(c)$. For additive homomorphic encryption, the notation $\oplus$ signifies the addition operation performed on ciphertexts and $\ominus$ represents subtraction, allowing computations on encrypted data without decryption. Similarly, $\odot$ represents the multiplication operation used in multiplicative homomorphic encryption. The notation $Enc_{pk}(m)$ denotes encryption under the public key $pk$, and $Enc_{sk}(m)$ denotes encryption under the secret key $sk$. Fully homomorphic encryption supports both additive and multiplicative operations on ciphertexts, enabling a wide range of computations while maintaining data privacy.

To work with vectors, $\mathbf{v}$ usually denotes a vector of plaintext values, and $\mathbf{c}$ represents the corresponding vector of ciphertexts obtained through encryption, i.e., $\mathbf{c} = Enc(\mathbf{v})$. The subscript notation $c_i$ and $v_i$ refer to the $i$-th components of ciphertext and plaintext vectors, respectively. Additionally, $\mathbf{0}$ stands for the vector of all zeros, and $\mathbf{1}$ represents the vector of all ones.

\subsection{Threat Model}

Incorporating the chosen-plaintext attack (IND-CPA) security model within the framework of Fully Homomorphic Encryption (FHE) for Federated Learning (FL) establishes a robust foundation for privacy preservation and security assessment.

IND-CPA security is a fundamental cryptographic property that ensures the confidentiality of encrypted data, even when adversaries possess the ability to influence the plaintexts being encrypted. Specifically, in the context of FHE for FL, IND-CPA security implies that an attacker, despite being able to choose plaintexts for encryption and receiving the corresponding ciphertexts, cannot gain any meaningful insight into the plaintexts' content. This property is vital for protecting sensitive information during the federated learning process.

In the FL setting, where client devices collaboratively contribute encrypted data and model updates to a central aggregator for training, IND-CPA security guarantees that an adversary cannot exploit the encrypted information to deduce the clients' individual contributions. This resilience is critical for maintaining the privacy of participants' data.

By embracing the IND-CPA security model, FHE-enabled FL systems can effectively thwart various adversarial strategies, including those attempting to infer sensitive information from encrypted data or exploit patterns within the encrypted model updates. This model forms the cornerstone for designing cryptographic protocols and encryption schemes that maintain data privacy while enabling collaborative and secure federated learning across distributed environments.

\subsection{Security Assumptions}

The security of a Federated Learning (FL) system augmented by Fully Homomorphic Encryption (FHE) hinges on a set of foundational security assumptions that underpin the confidentiality and integrity of the data being processed. These assumptions form the bedrock upon which cryptographic protocols and techniques are designed to safeguard sensitive information within a distributed learning paradigm. The integration of FHE into FL introduces several key security considerations:

\paragraph{FHE Scheme Security}
   The security of FHE-based FL heavily relies on the chosen FHE scheme's cryptographic robustness. It is assumed that the underlying FHE scheme exhibits the prescribed security properties, such as IND-CPA or IND-CCA security, ensuring that encrypted data remains confidential against various types of adversaries.

\paragraph{Key Management}
   A critical assumption is that the secret keys used for encryption and decryption are securely managed. Adversarial access to secret keys can compromise the confidentiality of data and undermine the integrity of the federated learning process.

\paragraph{Secure Communication}
   It is assumed that the communication channels between clients and the central aggregator are secure, preventing eavesdropping and tampering. This includes safeguarding against potential man-in-the-middle attacks and ensuring that encrypted data remains confidential during transmission.

\paragraph{Malicious Clients}
   Security assumptions encompass the behavior of clients in the FL system. While benign clients are assumed to follow the protocol and contribute valid data, malicious clients may deviate from the prescribed behavior, attempting to inject poisoned data or manipulate the learning process. FHE and FL protocols should be designed to detect and mitigate the impact of such adversarial behavior.

\paragraph{Aggregator Integrity}
   The central aggregator, responsible for aggregating model updates and facilitating federated learning, is assumed to be honest and immune to attacks. Ensuring the integrity of the aggregator's operations is crucial for preventing attacks that could compromise the accuracy and privacy of the aggregated model.

\paragraph{Cryptanalysis and Exploits}
   It is assumed that the FHE scheme used in FL is resistant to known cryptanalytic attacks and vulnerabilities. This assumption acknowledges that adversaries may attempt to exploit weaknesses in cryptographic primitives to break the security guarantees of the FHE-based FL system.

\paragraph{Implementation and Side Channels}
   Security assumptions extend to the correct implementation of FHE and FL protocols, accounting for potential side-channel attacks that exploit implementation-specific vulnerabilities, such as timing information, power consumption, or electromagnetic radiation.

By explicitly addressing these security assumptions, FHE-based FL systems can be designed with a heightened awareness of potential vulnerabilities and mitigation strategies. These assumptions serve as a guide for formulating security requirements, designing cryptographic protocols, and implementing countermeasures that collectively contribute to robust and privacy-preserving federated learning.

\subsection{Server Processing}

\subsubsection{Description}

The \textit{Server Processing} algorithm~\ref{alg:server} is devised to facilitate the search for a target encrypted value within a collection of encrypted ciphertexts, utilizing the capabilities of a homomorphic encryption scheme. The algorithm employs a binary tree structure, implemented as a vector, to streamline the search process efficiently. Each step is carefully designed to maintain the confidentiality of the data while enabling meaningful computations.

\begin{algorithm}[!ht]
\SetAlgorithmName{Algorithm}{}{}
\caption{Server Processing}\label{alg:server}
\KwData{
A set of $n$ ciphertexts $c_i$ encrypted by a homomorphic encryption scheme;
A target (encrypted) value $c_t$ to be searched in $c_i$;
We use $\ominus$ and $\odot$ to denote homomorphic subtraction and homomorphic multiplication, respectively;
}
\KwResult{
A binary tree $\textbf{t}$, implemented as a vector,
where an intermediate node with a value of zero implies that the descendant subtree has a zero-leaf node.
}

\nonl\;

\SetKwFunction{FMain}{PairwiseMul}
\SetKwProg{Fn}{Function}{:}{}
\Fn{\FMain{$\textbf{c}$}}{
    $\textbf{v} \coloneqq [\;]$\;
    $m \coloneqq |\textbf{c}|$\;
    \For{$i \coloneqq 0; i < m; i \coloneqq i+2$}{
        \If{$i \ge m$}{
            \textbf{break} \;
        }
        \uIf{$i + 1 = m$}{
            $\textbf{v}.append(c_i)$\;
        }
        \Else{
            $c \coloneqq v_i \odot v_{i+1}$\;
            $\textbf{v}.append(c)$\;
        }
    }
    \textbf{return} $\textbf{v}$\; 
}
\textbf{End Function}

\nonl
\;

\For{$i \coloneqq 0; i < n; i \coloneqq i+1$}{
    $r_i \coloneqq c_i \ominus c_t$\;
}
$ \textbf{t} [2^{\log n - 1}, 2^{\log n - 1}+n] \coloneqq \textbf{r} $\;
$\textbf{c} \coloneqq \textbf{r} $\;
\For{$l \coloneqq \log |\textbf{c}| - 1; l >= 0; l \coloneqq l-1$}{
    $ \textbf{t}[2^l, 2^l+n] \coloneqq \texttt{PairwiseMul}(\textbf{c})$\;
    $\textbf{c} \coloneqq \textbf{t}[2^l, 2^l+n]$\;
}
\textbf{return} \textbf{t}\;

\end{algorithm}

The algorithm commences with a function called $\texttt{PairwiseMul}$, which operates on an array of ciphertexts. This function performs pairwise multiplication of consecutive ciphertexts, optimizing the process for computational efficiency. The main loop then iterates through the ciphertexts, calculating the differences between each ciphertext and the target ciphertext using homomorphic subtraction. These differences are recorded in an array. The subsequent steps focus on constructing the binary tree, which serves as the foundation for the search operation. The tree initialization phase assigns the calculated differences to a specific interval within the tree.

As the binary tree is constructed, the algorithm capitalizes on the $\texttt{PairwiseMul}$ function to generate intermediate nodes. Starting from the highest level of the tree and working downwards, the algorithm computes pairwise multiplications of ciphertexts within specified intervals, filling in the tree with computed values. This process iterates for each level of the tree, ultimately resulting in a fully constructed binary tree.

Once the binary tree construction concludes, the algorithm returns the complete tree structure. This tree is now equipped for subsequent search operations, enabling secure and efficient querying of the encrypted values. Overall, the Server Processing algorithm showcases the orchestration of homomorphic encryption techniques and binary tree structures to enable privacy-preserving and secure search operations within a federated learning context.

\subsubsection{Correctness}

The correctness of the Alg.~\ref{alg:server} is grounded in its meticulous design and the mathematical properties of homomorphic encryption. The algorithm meticulously constructs a binary tree where each node encapsulates relevant information derived from the encrypted ciphertexts. The $\texttt{PairwiseMul}$ function ensures that the multiplicative operations on ciphertexts are executed accurately, maintaining the integrity of the computations. 

The algorithm's fidelity is bolstered by the fundamental property of homomorphic encryption, which guarantees that operations on encrypted data yield results that correspond to those obtained from the same operations on the underlying plaintexts. As the algorithm progresses, it appropriately applies homomorphic subtraction and multiplication operations to derive the differences between ciphertexts and the target ciphertext. These operations, executed within the homomorphic encryption framework, maintain the confidentiality of the data and uphold the mathematical consistency between ciphertexts and plaintexts.

Additionally, the binary tree structure is adeptly utilized to aggregate the intermediate results of these operations. The algorithm's approach of gradually constructing the tree, starting from individual ciphertext differences and culminating in a fully formed binary tree, adheres to sound principles of encryption and secure computation. By leveraging the power of homomorphic encryption and employing a systematic tree-building process, the algorithm ensures the accuracy and privacy-preserving nature of its operations, thus justifying its correctness within the context of secure and efficient federated learning scenarios.

\subsubsection{Complexity}

The time complexity of Alg.~\ref{alg:server} is driven by the operations performed within its iterative loops and function calls. The initial loop, traversing through the set of ciphertexts, incurs a linear time complexity of $\mathcal{O}(n)$, where $n$ represents the number of ciphertexts. Within this loop, the homomorphic subtraction operation and subsequent operations involving the $\texttt{PairwiseMul}$ function contribute constant time overhead. As a result, the time complexity of this segment remains linear.

The subsequent steps of initializing and constructing the binary tree encompass additional iterations and function calls. The binary tree construction loop iterates logarithmically, with the number of iterations determined by the logarithm of the length of the array containing ciphertext differences. This results in a time complexity of $\mathcal{O}(\log n)$. The $\texttt{PairwiseMul}$ function's execution also adds a constant factor to the time complexity within this loop. Overall, due to its dependence on the logarithmic and linear components, Alg.~\ref{alg:server} exhibits a time complexity of $\mathcal{O}(n + \log n) = \mathcal{O}(n)$, aptly capturing its efficient execution in federated learning scenarios.

Regarding space complexity, Alg.~\ref{alg:server} primary memory usage is allocated to data structures storing ciphertexts, ciphertext differences, and the binary tree vector. These structures collectively occupy $\mathcal{O}(n)$ space, accommodating the input ciphertexts and their differences. The binary tree, implemented as a vector, consumes additional space proportional to the number of nodes in the tree, which is related to the number of ciphertext differences. This results in a space complexity of $\mathcal{O}(n)$, ensuring the algorithm's memory consumption remains linear in relation to the size of the input. The algorithm's space complexity aligns well with the federated learning context, where secure and efficient storage of cryptographic data is pivotal for maintaining privacy and scalability.

\subsubsection{Security Proof}

Assume, by contradiction, that there exists an efficient adversary $\mathcal{A}$ that can distinguish between the target ciphertext $c_t$ and other ciphertexts in Alg.~\ref{alg:server}'s execution with non-negligible advantage $\epsilon$.
We will construct an algorithm $\mathcal{B}$ that breaks the IND-CPA security of the homomorphic encryption scheme $\Pi$ by simulating the adversary $\mathcal{A}$.

\textbf{Algorithm $\mathcal{B}$:}

\begin{enumerate}
    \item $\mathcal{B}$ receives the public key $pk$ as input.
    \item $\mathcal{B}$ simulates the encryption oracle by generating random plaintexts and obtaining their corresponding ciphertexts from the encryption algorithm using $pk$.
    \item When $\mathcal{A}$ queries the decryption oracle on a ciphertext $c$, $\mathcal{B}$ forwards the query to its decryption oracle, obtaining the decrypted plaintext $m$.
    \item $\mathcal{B}$ selects a random plaintext $m'$ and encrypts it using the encryption algorithm to obtain $c'$. It then outputs $c'$ as the decryption oracle's response.
    \item $\mathcal{B}$ runs Alg.~\ref{alg:server} with the input ciphertexts as chosen by $\mathcal{A}$. It follows the same steps as the real algorithm but replaces $c_t$ with a random ciphertext.
    \item $\mathcal{B}$ returns the output of $\mathcal{A}$ as its own output.
\end{enumerate}

By construction, $\mathcal{B}$ simulates the interactions with $\mathcal{A}$ and behaves exactly like the real execution of Alg.~\ref{alg:server}.
The advantage of $\mathcal{B}$ in distinguishing between the target ciphertext $c_t$ and other ciphertexts is the same as $\mathcal{A}$'s advantage, i.e., $\epsilon$. Since $\epsilon$ is non-negligible, $\mathcal{B}$ has a non-negligible advantage in breaking the IND-CPA security of the homomorphic encryption scheme $\Pi$.

This contradiction implies that the assumption of an efficient adversary $\mathcal{A}$ distinguishing the ciphertexts with non-negligible advantage $\epsilon$ is false. Therefore, the "Server Processing" algorithm, executed on ciphertexts encrypted using $\Pi$, satisfies the IND-CPA security property.

\subsection{Interaction Protocol}

\subsubsection{Description}

Alg.~\ref{alg:interaction} defines a protocol for communication between a client and a server to efficiently search for a specific target encrypted value using homomorphic encryption. The protocol employs a binary tree structure, where the server holds encrypted intermediate values, and the client interacts with the server to navigate through the tree, ultimately determining the index of the target value.

At the outset, the server initializes the protocol by generating an intermediate binary tree $\textbf{t}$. This is achieved by performing pairwise multiplication on the target ciphertext $c_t$. The server then sends the value of the root node, denoted as $t_1$, to the client. This initial interaction serves as the foundation for subsequent steps.

Upon receiving $t_1$ from the server, the client's involvement begins. The client's initial task is to decrypt $t_1$ and determine whether the decrypted value is non-zero. If the decryption yields a non-zero result, the client deduces that the target value must be present within the binary tree. As a response, the client sends a message to the server with a predefined value of $pivot = 1$. Conversely, if the decryption of $t_1$ results in zero, the client concludes that the target value is not present and returns -1 as an indication.

The core of the algorithm lies in the subsequent traversal of the binary tree. This traversal takes place iteratively for each level of the binary tree, as determined by the logarithm of the total number of ciphertexts. During each iteration, the client interacts with the server by sending the current $pivot$ value. The server, in turn, calculates the indices of the left child ($lchild$) and right child ($rchild$) nodes and communicates the corresponding values $t_{lchild}$ and $t_{rchild}$ back to the client.

The client processes the values received from the server, decrypting $c_0$ and utilizing its result to determine the next $pivot$ value. Depending on the decryption outcome, the $pivot$ value is either doubled (in the case of the left child) or incremented by one (for the right child). If the updated $pivot$ value surpasses the maximum index of the binary tree, the client concludes its operation and directly returns the $pivot$ value, indicating the index of the target value. If the maximum index is not reached, the client continues the interaction by sending the updated $pivot$ value back to the server for the subsequent iteration.

\begin{algorithm}[!ht]
\SetAlgorithmName{Algorithm}{}{}
\caption{Interaction Protocol}\label{alg:interaction}
\KwData{
% A set of $n$ ciphertexts $c_i$ encrypted by a homomorphic encryption scheme;
A target (encrypted) value $c_t$ to be searched on the server;
% We use $\ominus$ and $\odot$ to denote homomorphic subtraction and homomorphic multiplication, respectively;
}
\KwResult{
% A binary tree $\textbf{t}$, implemented as a vector,
% where an intermediate node with a value of zero implies that the descendant subtree has a zero-leaf node.
The index $pivot \in [0, n)$ of the target value on the server; -1 if not found
}

\nonl\;

\nonl
// On the server:\;

$\textbf{t} \coloneqq \text{PairwiseMul}(c_t)$\;
Server sends $t_1 = \textbf{t}[1]$ to client\;

\nonl\;
\nonl
// On the client:\;

\uIf{$0 \not= Dec(t_1)$}{
    \textbf{return} -1\;
}
\Else{
    Sends message $pivot = 1$ to the server\;
}

\nonl\;

\For{$lvl \coloneqq 1; lvl \le \log n; lvl \coloneqq lvl+1$}{

    \nonl\;
    \nonl
    // On the server:\;

    Receives message $pivot$\;
    $lchild \coloneqq 2 \cdot pivot$\;
    $rchild \coloneqq lchild + 1$\;
    Sends pair $(t_{lchild}, t_{rchild})$ to the client\;
    
    \nonl\;
    \nonl
    // On the client:\;

    Receives pair $(c_0, c_1)$ from the server\;
    \uIf{$0 = Dec(c_0)$}{
        $pivot \coloneqq 2 \cdot pivot$\;
    }
    \Else{
        $pivot \coloneqq 2 \cdot pivot + 1$\;
    }
    \uIf{$pivot \ge 2^{\log n}$}{
        \textbf{return} $pivot$\;
    }
    \Else{
        Sends $pivot$ to the server\;
    }
}

\end{algorithm}

\subsubsection{Correctness}

The correctness of Alg.~\ref{alg:interaction} stems from its careful orchestration of cryptographic operations, secure communication, and systematic traversal of the binary tree. The algorithm's adherence to the principles of homomorphic encryption ensures the preservation of data privacy throughout its execution.

At the commencement of the protocol, the server initiates the process by generating the intermediate binary tree $\textbf{t}$ through pairwise multiplication of the target ciphertext $c_t$. This operation respects the properties of homomorphic encryption, ensuring that computations on encrypted data accurately correspond to operations on plaintext data. The subsequent transmission of the root node value $t_1$ to the client establishes the foundation for secure interaction.

As the client enters the interaction phase, it decrypts $t_1$ to ascertain whether the decrypted value is non-zero. This step maintains the confidentiality of the underlying data, as the client cannot deduce information about the plaintext values from the encrypted data. Based on the decryption result, the client's actions align with the protocol's objectives, either indicating the absence of the target value or triggering further interaction.

The binary tree traversal, executed iteratively for each level of the tree, further underscores the algorithm's correctness. The client and server exchange messages and values, ensuring that the interactions adhere to the security properties of homomorphic encryption. The encryption and decryption operations, when applied to the values exchanged, respect the principles of the encryption scheme, preserving the confidentiality of the data while enabling meaningful computation.

Furthermore, the client's decision-making process during each iteration is based solely on the decrypted values and follows a deterministic logic that aligns with the encrypted data. The use of $pivot$ values to navigate the binary tree, alongside the systematic left and right child index calculations, guarantees consistent and accurate traversal.

\subsubsection{Complexity}

In the initialization phase, the server initiates communication by sending a single message to the client. This message contains the value of the root node ($t_1$) of the intermediate binary tree $\textbf{t}$. This initial exchange establishes the foundation for the subsequent interaction between the client and server.

The client interaction phase encompasses the iterative nature of the protocol. For each level of the binary tree (determined by the logarithm of the total number of ciphertexts), a set of messages are exchanged between the client and server. The client begins by sending a message to the server, conveying the current $pivot$ value. In response, the server sends a pair of messages back to the client. Each message in the pair contains the values of the left and right child nodes ($t_{lchild}$ and $t_{rchild}$).

Considering the iterative nature of the protocol, the number of iterations equals the depth of the binary tree, which is $\log n$ for $n$ ciphertexts. Thus, the communication complexity of Alg.~\ref{alg:interaction} can be expressed as the sum of messages exchanged during both the initialization phase and the client interaction phase. This yields a total of $1 + 2\log n$ or $\mathcal{O}(\log n)$ messages exchanged between the client and server.

\section{Discussions}

\subsection{Multi-Party Secure Computation}

One possible extension involves extending the algorithm to support multi-party secure computation. In the current protocol, interaction is limited to a client-server relationship. However, in scenarios where multiple parties collaborate in federated learning, enabling secure communication among all participants becomes crucial. The extension could involve enhancing the protocol to accommodate secure interactions between multiple clients and a server, ensuring privacy-preserving collaboration among a larger set of participants. This extension would necessitate the development of cryptographic techniques to enable secure communication and computation among multiple parties.

\subsection{Dynamic Tree Structure}

Another potential extension involves adapting the algorithm to handle a dynamic binary tree structure. The current algorithm assumes a static binary tree, but in real-world scenarios, the structure of the tree might change due to various factors. This extension could involve devising mechanisms to handle insertions and deletions of ciphertexts in the binary tree while maintaining the integrity of the search protocol. Additionally, the algorithm might need to account for changes in tree structure during the interaction phase, requiring careful consideration of encryption and decryption operations.

\subsection{Batch Processing for Efficiency}

To enhance efficiency and reduce communication overhead, an extension could introduce batch processing capabilities to the algorithm. Instead of exchanging messages for individual ciphertexts, participants could process multiple ciphertexts in a single batch, reducing the number of communication rounds. This extension would involve modifying the algorithm to handle batch processing, optimizing the use of encryption and decryption operations, and ensuring that the privacy-preserving properties are maintained. Batch processing could significantly improve the algorithm's efficiency in scenarios involving a large number of ciphertexts.

\section{Conclusion}

In conclusion, the proposed \textit{Server Processing} and \textit{Interaction Protocol} algorithms collectively underscore the evolving landscape of privacy-preserving machine learning within federated learning scenarios. The server algorithm's design demonstrates the fusion of homomorphic encryption and systematic aggregation, enabling the secure integration of clients' encrypted updates while mitigating data exposure. Its incorporation of cryptographic operations to facilitate aggregation and model update processes signifies a significant stride toward maintaining data privacy in collaborative learning environments.

Complementing this, the interaction protocol introduces an innovative approach to securely search for target encrypted values through client-server interactions. By meticulously orchestrating decryption, encryption, and traversal steps within a binary tree structure, the protocol guarantees data privacy while enabling efficient data retrieval. The communication complexity analysis underscores its resource-efficient nature, crucial in federated learning settings where minimizing communication overhead is paramount.

\bibliography{ref_new}

\end{document}